\documentclass[lettersize,journal]{IEEEtran}
\usepackage{cite}
\usepackage{amsmath,amssymb,amsfonts}
\usepackage{graphicx}
\usepackage{textcomp}
\usepackage{xcolor}
\usepackage{subcaption}

\usepackage{booktabs}
\usepackage{bm}
\usepackage{url}
\usepackage[ruled,linesnumbered]{algorithm2e}

\usepackage{amsmath,amsfonts}
\usepackage{array}
\usepackage[caption=false,font=normalsize,labelfont=sf,textfont=sf]{subfig}
\usepackage{textcomp}
\usepackage{stfloats}
\usepackage{url}
\usepackage{verbatim}
\usepackage{graphicx}
\def\BibTeX{{\rm B\kern-.05em{\sc i\kern-.025em b}\kern-.08em
    T\kern-.1667em\lower.7ex\hbox{E}\kern-.125emX}}
\usepackage{balance}

\begin{document}
\title{PFCM: Poisson flow consistency models for low-dose CT image denoising}
\author{Dennis Hein, Grant Stevens, $\text{Adam Wang}^\dagger$, and $\text{Ge Wang}^\dagger$
\thanks{D. H. is with the Department of Radiology, Stanford University, Stanford, CA, USA (\textit{heind@stanford.edu}), the Department of Physics, KTH Royal Institute of Technology, Stockholm, Sweden, and MedTechLabs, Karolinska University Hospital, Stockholm, Sweden. G.S. is with GE HealthCare, Waukesha, WI USA. A. W. is with the Department of Radiology and the Department of Electrical Engineering, Stanford University, Stanford, CA, USA. G. W. is with the Department of Biomedical Engineering, School of Engineering, Biomedical Imaging Center, Center for Biotechnology and Interdisciplinary Studies, Rensselaer Polytechnic Institute, Troy, NY, USA. $^\dagger$Joint senior authors.}}

\maketitle

\begin{abstract}
X-ray computed tomography (CT) is widely used for medical diagnosis and treatment planning; however, concerns about ionizing radiation exposure drive efforts to optimize image quality at lower doses. This study introduces Poisson Flow Consistency Models (PFCM), a novel family of deep generative models that combines the robustness of PFGM++ with the efficient single-step sampling of consistency models. PFCM are derived by generalizing consistency distillation to PFGM++ through a change-of-variables and an updated noise distribution. As a distilled version of PFGM++, PFCM inherit the ability to trade off robustness for rigidity via the hyperparameter $D \in (0,\infty)$. A fact that we exploit to adapt this novel generative model for the task of low-dose CT image denoising, via a ``task-specific'' sampler that ``hijacks'' the generative process by replacing an intermediate state with the low-dose CT image. While this ``hijacking'' introduces a severe mismatch---the noise characteristics of low-dose CT images are different from that of intermediate states in the Poisson flow process---we show that the inherent robustness of PFCM at small $D$ effectively mitigates this issue. The resulting sampler achieves excellent performance in terms of LPIPS, SSIM, and PSNR on the Mayo low-dose CT dataset. By contrast, an analogous sampler based on standard consistency models is found to be significantly less robust under the same conditions, highlighting the importance of a tunable $D$ afforded by our novel framework. To highlight generalizability, we show effective denoising of clinical images from a prototype photon-counting system reconstructed using a sharper kernel and at a range of energy levels.
\end{abstract}


\IEEEPARstart{X}{-ray} computed tomography (CT) has become an indispensable tool in modern medicine, playing a crucial role in the diagnosis and treatment planning for a diverse array of conditions, including stroke, cancer, and cardiovascular disease. However, the medical community remains vigilant about the potential risks associated with ionizing radiation, even at low doses~\cite{degonzales2004, brenner2007}. Consequently, there is an ongoing effort within the field to optimize CT protocols and develop effective denoising techniques, aiming to minimize radiation exposure while preserving the high diagnostic quality essential for accurate clinical assessment~\cite{wang2020,koetzier2023}. Photon-counting CT (PCCT), the latest advancement in CT detector technology, mitigates the influence of electronic noise, thereby enabling enhanced low-dose imaging. Additionally, PCCT offers substantial improvements in spatial and energy resolution~\cite{willemink2018,flohr2020,danielsson2021,hsieh2021,higashigaito2022,rajendran2022}. As higher spatial or energy resolution shrinks the photon count per voxel or energy bin, and thus elevates noise, there is a need for robust denoising methods to realize the full potential of PCCT systems. 

Techniques used for CT noise suppression can roughly be categorized as: iterative reconstruction~\cite{wang2006,thibault2007,sidky2008,yu2009,tian2011,stayman2013,zeng2017}, pre- and post-processing~\cite{lariviere2005,ma2011,zhang2016,chen2017,wolterink2017,yang2018, shan2019, kim2019, kim2020, yuan2020, makinen2020, li2021, wang2023, niu2023, liu2023, tivnan2023, hein2024, hein2024b}, and dual-domain methods\cite{niu2022}. Although each of these approaches comes with its own advantages and disadvantages, deep learning-based post-processing techniques have become increasingly popular due to remarkable effectiveness of reducing noise without evident over-smoothing often outperforming traditional methods in terms of both speed and accuracy.

Diffusion and Poisson flow models have demonstrated remarkable success for a range of generative and image processing tasks, both in the unconditional~\cite{sohl-dickstein2015,ho2020,nichol2021,song2021,song2021b,karras2022,xu2022,xu2023} and conditional~\cite{song2021, batzolis2021,chung2021,song2021c,saharia2022b,saharia2023,chung2023,liu2023, tivnan2023, ge2023, hein2024, hein2024b} setting. Poisson flow generative models (PFGM)++~\cite{xu2023} look to electrostatics for inspiration, instead of non-equilibrium thermodynamics as is the case for diffusion models, and treat the $N$-dimensional data (where $N:= n\times n$) as electric charges in a $(N+D)$-dimensional augmented space. In this framework, electric field lines define trajectories via ordinary differential equations (ODEs), establishing a bijective mapping between the data distribution of interest and an easy-to-sample noise distribution. A key feature of PFGM++ is its contribution to unifying physics-inspired generative models. The authors prove that as $D\rightarrow \infty$ and $r = \sigma \sqrt{D}$, the training and sampling processes of PFGM++ converge to those of Elucidated Diffusion Models (EDM)~\cite{karras2022}. This convergence property endows PFGM++ with remarkable flexibility: by tuning the hyperparameter $D$, users can balance robustness and rigidity while maintaining diffusion models as a special case. Moreover, Xu et al.~\cite{xu2023} demonstrate that existing training and sampling algorithms for EDM can be directly adapted for PFGM++. This adaptation requires only a simple change-of-variables and an updated noise distribution, facilitating efficient knowledge transfer between these related frameworks.

Diffusion models, while powerful, rely on iterative sampling. This process allows for zero-shot data editing and quality-compute trade-offs but is computationally expensive, limiting real-time applications. Consistency models~\cite{song2023,song2023b} are a nascent class of deep generative models, built upon the continuous-time diffusion models framework, that overcome this challenge by learning a function that directly maps noise to data. This approach enables high-quality single-step sampling without adversarial training. The fundamental principle underlying consistency models is self-consistency. This principle ensures that any two points on the same solution trajectory map to the same initial point. Notably, consistency models retain the ability for zero-shot data editing and quality-compute trade-offs. However, in its single-step formulation, consistency models are considerably harder to train than their more forgiving, multi-step, diffusion models counterparts~\cite{heek2024}. 

\begin{figure*}
    \centering
    \includegraphics[width=0.75\textwidth]{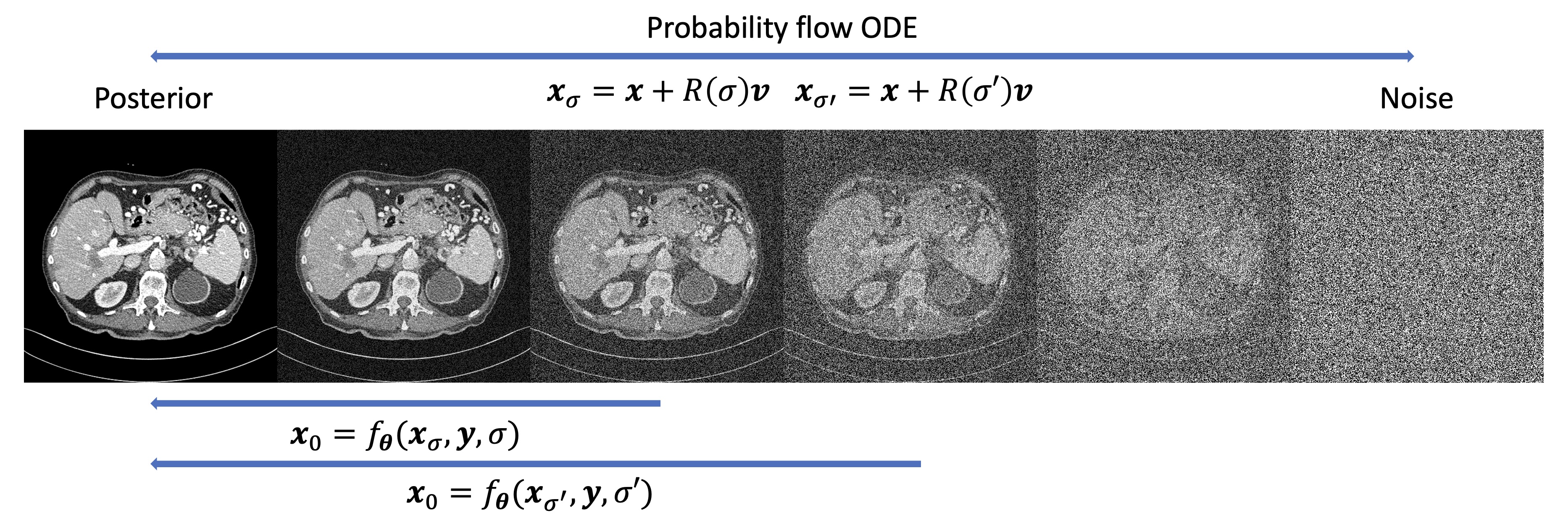}
    \caption{Overview of PFCM for the task of low-dose CT image denoising. The objective is to obtain a high-quality reconstruction $\bm{\hat{x}} \in \mathbb{R}^N$ of the normal dose CT image $\bm{x} \in \mathbb{R}^N$ based on the low-dose CT image $\bm{y}\in \mathbb{R}^N$. Approaching this as a statistical inverse problem, our solution is a sample $\hat{\bm{x}}$ from the posterior distribution $p(\bm{x}|\bm{y}).$ We enable sampling from said posterior by first training PFGM++~\cite{xu2023} in a supervised fashion to directly learn a mapping between the prior noise distribution and the posterior distribution of interest via the probability flow ODE by feeding the noisy image, $\bm{y}$, as additional input at training and test time. We subsequently
    obtain Poisson flow consistency models (PFCM) via consistency distillation~\cite{song2023} using a change-of-variables and an updated noise distribution. A sample $\hat{\bm{x}}\sim p(\bm{x}|\bm{y})$ can then be obtained in a single-step as $\hat{\bm{x}}=f_{\bm{\theta}}(\bm{x}_\sigma,\bm{y},\sigma),$ for any intermediate $(\bm{x}_\sigma,\sigma)$, including for $\sigma_{\text{max}},$ where $\bm{x}_{\sigma_{\text{max}}}$ is pure noise. To adapt this setup for low-dose CT image denoising, where data fidelity is of utmost importance, we propose a ``task-specific'' sampler, which ``hijacks'' the sampling process, replacing the intermediate state $\bm{x}_{\sigma}$ with the low-dose CT image $\bm{y}$. Crucial to this approach is the robustness to ``missteps'' in the sampling process, afforded by PFCM via tunable $D \in (0,\infty)$.}
    \label{pfcm_process}
\end{figure*}

In this paper, we introduce a novel family of deep generative models that combines the robustness of tunable $D\in (0,\infty)$ in PFGM++~\cite{xu2023} with the high-quality single-step sampling of consistency models~\cite{song2023}. We demonstrate this framework's effectiveness for the inverse problem of low-dose CT image denoising. Our main contributions are: 1) We extend the concept of distilling pre-trained diffusion models into single-step samplers via consistency distillation to PFGM++. This generalization is achieved through a simple change-of-variables and an updated noise distribution. We term this novel framework Poisson Flow Consistency Models (PFCM) and demonstrate its efficacy as a posterior sampler for low-dose CT image denoising. Our approach is illustrated in Fig.~\ref{pfcm_process}. 2) We show that there exists a $D \in (0,\infty)$ for which PFCM outperforms consistency models for the low-dose CT denoising task. 3) Addressing the issue of excessive stochastic variation in images produced by both consistency models and PFCM, we propose a task-specific sampler, effectively adapting this framework to the medical imaging domain. This sampler leverages the flexibility of the PFCM framework by setting $D$ relatively small, resulting in a highly robust model. Due to its robustness, we can manipulate the sampling process by further conditioning on the low-dose CT image via ``hijacking'', leading to greatly improved data fidelity. Notably, the same framework using consistency models fails. 4) We demonstrate our method's generalizability by effectively denoising virtual monoenergetic images from a prototype photon-counting CT (PCCT) system reconstructed at various energy levels and with a sharper kernel. 
%Code for this project is available at: \url{https://github.com/dennishein/pfcm}. 

\section{Background}
\subsection{Problem formulation}
Our objective is to obtain a high-quality reconstruction $\bm{\hat{x}} \in \mathbb{R}^N$ of the true image $\bm{x} \in \mathbb{R}^N$ based on noisy observations $\bm{y} = \mathcal{F}(\bm{x}) \in \mathbb{R}^N$. Here, $\mathcal{F}: \mathbb{R}^N \rightarrow \mathbb{R}^N$ denotes a noise degradation operator that encompasses various factors, including quantum noise~\cite{chen2017}. For notational brevity, we define $N := n \times n$, where $n$ represents the image resolution. We approach this as a statistical inverse problem. Given the prior distribution $\bm{x} \sim p(\bm{x})$, the solution is obtained by sampling $\hat{\bm{x}}$ from the posterior distribution $p(\bm{x}|\bm{y})$. We will train PFCM in a supervised fashion to directly learn the trajectory between the easy-to-sample noise distribution and the posterior distribution of interest by feeding the noisy image $\bm{y}$ as additional input. 

\subsection{Consistency models}
Consistency models (CM)~\cite{song2023,song2023b} build on the continuous-time diffusion models~\cite{song2021,karras2022}, which extend the idea of learning to model complex data distribution via a sequence of simpler denoising steps from DDPM~\cite{sohl-dickstein2015,ho2020} to the limit where the number of noise scales becomes infinite. In this setting, the data distribution, $p_{\text{data}}$, is perturbed into a noise distribution via a stochastic differential equation (SDE). Remarkably, this SDE has a corresponding ordinary differential equation (ODE) called the probability flow (PF) ODE~\cite{song2021}. Using the settings in EDM~\cite{karras2022} the probability flow ODE takes the form 
\begin{equation}
    d\bm{x} = -\sigma \nabla_{\bm{x}} \log p_{\sigma}(\bm{x}) dt,
    \label{ode_simple}
\end{equation}
for $\sigma \in [\sigma_{\text{min}},\sigma_{\text{{max}}}]$, where $\nabla_{\bm{{x}}} \log p_{\sigma}(\bm{x})$ is the score function of the perturbed data distribution $p_{\sigma}(\bm{x})$, and $\sigma(t)$ is the predefined noise scales. $\sigma_{\text{min}}$ is chosen sufficiently small, but not zero for numerical stability, such that $p_{\sigma_{\text{min}}}(\bm{x}) \approx p_{\text{data}}(\bm{x})$ and $\sigma_{\text{max}}$ sufficiently large such that $p_{\sigma_{\text{max}}}(\bm{x}) \approx \pi(\bm{x})$, some unstructured, easy-to-sample, noise distribution. In this case $\pi(\bm{x}):=\mathcal{N}(\bm{0},\sigma_{\text{max}}\bm{I}).$ 

In particular, for the solution trajectory $\{\bm{x}_\sigma\}_{\sigma\in[\sigma_{\text{min}},\sigma_{\text{max}}]}$, CM aim to learn a consistency function
\begin{equation}
f (\bm{x}_\sigma,\sigma) \mapsto \bm{x}_{\sigma_{\text{min}}}
\end{equation}
by enforcing self-consistency
\begin{equation}
f(\bm{x}_\sigma,\sigma)=f(\bm{x}_{\sigma'},\sigma'), \quad \forall \sigma,\sigma' \in [\sigma_{\text{min}},\sigma_{\text{max}}].
\end{equation}
Let $F_{\bm{\theta}}(\bm{x},t)$ denote a free-form deep neural network. The desired solution is then enforced via the boundary condition
\begin{equation}
f_{\bm{\theta}}(\bm{x},\sigma)=c_{\text{skip}}(\sigma)\bm{x}+c_{\text{out}}(\sigma) F_{\bm{\theta}}(\bm{x},\sigma)
\end{equation}
where $c_{\text{skip}}(\sigma)$ and $c_{\text{out}}(\sigma)$ are differentiable functions such that $c_{\text{skip}}(\sigma_{\text{min}})=1$ and $c_{\text{out}}(\sigma_{\text{min}})=0$, and $f_{\bm{\theta}}(\bm{x},\sigma)$ represents the deep neural network used to approximate the consistency function.

To learn the consistency function, the domain $[\sigma_{\text{min}},\sigma_{\text{max}}]$ is discretized using a sequence of noise scales $\sigma_{\text{min}}=\sigma_1 < \sigma_2 < \cdots < \sigma_N = \sigma_{\text{max}}$. CM are then trained using the consistency matching loss
\begin{equation}
\mathbb{E} \left[\lambda(\sigma_i) d(f_{\bm{\theta}}(\bm{x}_{\sigma_{i+1}},\sigma_{i+1}),f_{\bm{\theta}^-}(\breve{\bm{x}}_{\sigma_i},\sigma_{i}))\right],
\label{cm_loss}
\end{equation}
where
\begin{equation}
\breve{\bm{x}}_{\sigma_{i}}=\bm{x}_{\sigma_{i+1}}-(\sigma_i-\sigma_{i+1})\sigma_{i+1}\nabla_{\bm{x}} \log p_{\sigma_{i+1}}(\bm{x})|_{\bm{x}=\bm{x}_{\sigma_{i+1}}},
\label{ode_solve}
\end{equation}
is a single-step in the reverse direction solving the probability flow ODE in Eq.~\eqref{ode_simple}. Here, $\lambda(\cdot)\in \mathbb{R}_+$ is a positive weighting function, and $d(\cdot,\cdot)$ is a metric function, for instance Learned Perceptual Image Patch Similarity (LPIPS)~\cite{zhang2018}. The expectation in Eq. \eqref{cm_loss} is taken with respect to $i\sim \mathcal{U}[1,N-1]$, a uniform distribution over integers $i=1,\ldots,N-1$, and $\bm{x}_{\sigma_{i+1}} \sim p_{\sigma_{i+1}}(\bm{x})$. The objective in Eq.~\eqref{cm_loss} is minimized via stochastic gradient descent on parameters $\bm{\theta}$ whilst $\bm{\theta}^-$ are updated via the exponential moving average (EMA): $\bm{\theta^-} \leftarrow \text{stopgrad}(\mu\bm{\theta^-}+(1-\mu)\bm{\theta}),$ where $0\leq \mu \leq 1$ is the decay rate.

Directly minimizing the objective in Eq.~\eqref{ode_solve} is not feasible in practice since $\nabla_{\bm{x}}\log p_{\sigma_{i+1}}(\bm{x})$ is unknown. Song et al.~\cite{song2023} present two algorithms to circumvent this issue: consistency distillation and consistency training.
Consistency distillation relies on a pre-trained score network, $s_{\bm{\phi}}(\bm{x},\sigma)$, to estimate the score function, $\nabla_{\bm{x}}\log p_{\sigma_{i+1}}(\bm{x})$. The pre-trained score network is used instead of $\nabla_{\bm{x}}\log p_{\sigma_{i+1}}(\bm{x})$ in Eq.~\eqref{ode_solve} to generate a pair of adjacent points $(\breve{\bm{x}}_{\sigma_{i}},\bm{x}_{\sigma_{i+1}})$ on the solution trajectory.
Consistency training, on the other hand, is a stand-alone model using the Gaussian perturbation kernel to approximate two adjacent points on the solution trajectory. 

Once an approximation of the consistency function is obtained, a sample can be generated by drawing an initial noise vector from the noise distribution and passing it through the consistency model: $\hat{\bm{x}}=f_{\bm{\theta}}(\bm{z},\sigma_{\text{max}})$.

\subsection{PFGM++}
PFGM++~\cite{xu2023} is closely related to continuous-time diffusion models~\cite{song2021,karras2022} in theory and in practice. The training and sampling algorithms from Karras et al.~\cite{karras2022} can be applied directly via a simple change-of-variables and an updated noise distribution~\cite{xu2023}. It is also possible to show that the training and sampling processes of PFGM++ will converge to that of diffusion models~\cite{karras2022} is the $D\rightarrow \infty, r=\sigma \sqrt{D},$ limit. Inspired by electrostatics, PFGM++ treats the $N$-dimensional data as an electric charge in a $N+D$-dimensional augmented space. Let $\tilde{\bm{x}}:=(\bm{x},\bm{0}) \in \mathbb{R}^{N+D}$ and $\tilde{\bm{x}}_\sigma:=(\bm{x}_\sigma,\bm{z}) \in \mathbb{R}^{N+D}$ denote the augmented ground truth and perturbed data, respectively. The objective of interest is then the high dimensional electric field 
\begin{equation}
    \bm{E} (\bm{\tilde{x}}_\sigma) = \frac{1}{S_{N+D-1}(1)} \int \frac{\bm{\tilde{x}}_\sigma-\bm{\tilde{x}}}{||\bm{\tilde{x}}_\sigma-\bm{\tilde{x}}||^{N+D}} p(\bm{x}) d\bm{x},
    \label{pfgmpp_E}
\end{equation}
where $S_{N+D-1}(1)$ is the surface area of the unit $(N+D-1)$-sphere, $p(\bm{x})$ is the distribution of the ground truth data. Notably, this electric field is rotationally symmetric on the $D$-dimensional cylinder $\sum_{i=1}^D z_i^2 = r^2, \forall r > 0$ and thus it suffices to track the norm of the augmented variables $r = r(\bm{\tilde{x}}_\sigma):=||\bm{z}||_2.$ For notational brevity, redefine $\tilde{\bm{x}}:=(\bm{x},0) \in \mathbb{R}^{N+1}$ and $\tilde{\bm{x}}_\sigma:=(\bm{x}_\sigma,r) \in \mathbb{R}^{N+1}.$ This dimensionality reduction, afforded by symmetry, yields the following ODE 
\begin{equation}
    d\bm{x}_\sigma = \bm{E} (\bm{\tilde{x}}_\sigma)_{\bm{x}_\sigma} \cdot E(\bm{\tilde{x}}_\sigma)_r^{-1} \label{pfgmpp_ode} dr, 
\end{equation}
where $\bm{E}(\bm{\tilde{x}}_\sigma)_{\bm{x}_\sigma}=\frac{1}{S_{N+D-1}(1)} \int \frac{\bm{x}_\sigma-\bm{x}}{||\bm{\tilde{x}}_\sigma-\bm{\tilde{x}}||^{N+D}}p(\bm{x})d\bm{x}$ and $E(\bm{\tilde{x}}_\sigma)_r=\frac{1}{S_{N+D-1}(1)} \int \frac{r}{||\bm{\tilde{x}}_\sigma-\bm{\tilde{x}}||^{N+D}}p(\bm{x})d\bm{x}$ are the $\bm{x}_\sigma$ and $r$ components of Eq.~\eqref{pfgmpp_E}, respectively. Eq.~\eqref{pfgmpp_ode} defines a bijective (invertible) mapping between the ground truth data on the $r=0$ $(\bm{z}=\bm{0})$ hyperplane and an easy-to-sample prior noise distribution on the $r=r_{\max}$ hyper-cylinder~\cite{xu2023}. As is the case for diffusion models~\cite{song2021,karras2022}, PFGM++ uses a perturbation based objective function. Let $p_r(\bm{x}_\sigma | \bm{x})$ denote the perturbation kernel, which samples perturbed $\bm{x}_\sigma$ from ground truth $\bm{x}$, and $p(r)$ the training distribution over $r.$ Then the objective of interest is 
\begin{equation}
    \mathbb{E}_{r\sim p(r)} \mathbb{E}_{\bm{x}\sim p(\bm{x})} \mathbb{E}_{\bm{x}_\sigma \sim p_r(\bm{x}_\sigma|\bm{x})} \left[ ||s_\phi (\bm{\tilde{x}}_\sigma)-\frac{\bm{x}_\sigma-\bm{x}}{r/\sqrt{D}} ||_2^2\right] 
    \label{pfgmpp_obj_final}
\end{equation}
We can ensure that the minimizer of Eq.~\eqref{pfgmpp_obj_final} is $s^*_\phi(\bm{\tilde{x}}_\sigma) = \sqrt{D} \bm{E} (\bm{\tilde{x}}_\sigma)_{\bm{x}_\sigma} \cdot E(\bm{\tilde{x}}_ \sigma)_r^{-1}$ by choosing the perturbation kernel to be $p_r(\bm{x}_\sigma|\bm{x}) \propto 1/(|| \bm{x}_\sigma-\bm{x}||_2^2+r^2)^{\frac{N+D}{2}}.$ As with diffusion models, a sample can then be generated by solving $d\bm{x}_\sigma/dr = \bm{E}(\bm{\tilde{x}}_\sigma)_{\bm{x}_\sigma}/E(\bm{\tilde{x}}_\sigma)_r = s^*_\phi(\bm{\tilde{x}}_\sigma)/\sqrt{D}.$

\section{Main contributions}
\subsection{PFCM}
For simplicity, we will only treat PFCM via consistency distillation and leave consistency training to future work. For consistency distillation, the probability flow ODE solver from the pre-trained diffusion model is applied as is, including for the second-order Heun solver proposed in Karras et al.~\cite{karras2022}. In addition, as shown in PFGM++~\cite{xu2023}, this solver can be re-purposed for PFGM++ via a simple change-of-variables. Hence, using the noise distribution from PFGM++ we can sample $\bm{x}_{\sigma_{i+1}}$ from the PFGM++ solution trajectory and, as with CM, obtain the corresponding $\breve{\bm{x}}_{\sigma_{i}}$ by taking a single reverse step of the probability flow ODE solver using the pre-trained network from PFGM++, which estimates $\sqrt{D} \bm{E} (\bm{\tilde{x}}_\sigma)_{\bm{x}_\sigma} \cdot E(\bm{\tilde{x}}_ \sigma)_r^{-1}.$ In particular, for our discretized domain $[\sigma_{\text{min}},\sigma_{\text{{max}}}]$, let $\Phi(\cdot,\cdot, \bm{\phi})$ denote the update function of a one-step ODE solver applied to the probability flow ODE. We can then write Eq. \eqref{ode_solve} as 
\begin{equation}
    \breve{\bm{x}}_{\sigma_{i}}=\bm{x}_{\sigma_{i+1}}+(\sigma_{i}-\sigma_{i+1})\Phi(\bm{x}_{\sigma_{i+1}},\sigma_{i+1}, \bm{\phi}),
    \label{ode_solve_gen}
\end{equation}
where $\Phi(\bm{x}_{\sigma_{i+1}},\sigma_{i+1}, \bm{\phi}):=-\sigma_{i+1}s_{\bm{\phi}}(\bm{x}_{\sigma_{i+1}},\sigma_{i+1})$, using the first order Euler step for simplicity. More generally, Eq.~\eqref{ode_solve_gen} is taking a single-step in the reverse direction solving a probability flow ODE on the form 
\begin{equation}
    d\bm{x} = \Phi(\bm{x},\sigma, \bm{\phi}) dt. 
    \label{gen_ode}
\end{equation}
Using the fact that $\sigma(t)=t,$ $\bm{\tilde{x}}_\sigma:=(\bm{x}_\sigma,r),$ and the alignment formula $r=\sigma \sqrt{D}$~\cite{xu2023}, we can do the change-of-variables $dr=d\sigma \sqrt{D}=dt \sqrt{D}$ and rewrite $d\bm{x}/dr$ in Eq. \eqref{pfgmpp_ode} as 
\begin{equation}
    d\bm{x} = s_{\bm{\phi}}(\bm{\tilde{x}})/\sqrt{D} dr = s_{\bm{\phi}}(\bm{\tilde{x}})dt.
\end{equation}
Hence, we can write the probability flow ODE in Eq.~\eqref{pfgmpp_ode} on the form in Eq.~\eqref{gen_ode} with $\Phi_{\text{PFGM++}}(\bm{x},\sigma, \bm{\phi}):=s_{\bm{\phi}}(\tilde{\bm{x}})=\sqrt{D} \bm{E} (\bm{\tilde{x}}_\sigma)_{\bm{x}_\sigma} \cdot E(\bm{\tilde{x}}_ \sigma)_r^{-1}.$ PFCM can then be trained via consistency distillation by using $\Phi_{\text{PFGM++}}(\cdot,\cdot,\bm{\phi})$, from a pre-trained PFGM++, and an updated noise distribution.

As for diffusion models, $p_r(\cdot|\bm{x})$ enables sampling of perturbed $\bm{x}_\sigma$ from ground truth $\bm{x}.$ However, sampling from $p_r(\cdot|\bm{x})$ is slightly more involved than from a simple Gaussian. In particular, $p_r(\cdot|\bm{x})$ is first split into hyperspherical coordinates to $\mathcal{U}_\psi(\psi)p_r(R)$, where $\mathcal{U}_\psi$ denotes the uniform distribution of the angle component and the distribution of $R=||\bm{x}_\sigma-\bm{x}||$, the perturbed radius, is $p_r(R)\propto R^{N-1}/(R^2+r^2)^{\frac{N+D}{2}}.$
In practice, Xu et al.~\cite{xu2023} propose to sample from $p_r(\cdot|\bm{x})$ through the following steps: 1) get $r_i$ from $\sigma_i$ via the alignment formula $r_{i}=\sigma_{i} \sqrt{D}$. 2) sample radii $R_{i}$ from $p_{r_{i}}(R)$, 3) sample uniform angles $\bm{v}_{i}\leftarrow\bm{u}_{i}/||\bm{u}_{i}||_2,$ where $\bm{u}_{i}\sim \mathcal{N}(\bm{0};\bm{I})$, and 4) get perturbed data $\bm{x}_{\sigma_{i}} = \bm{x}+R_{i}\bm{v}_{i} \sim p_r(\bm{x}_{\sigma_i}|\bm{x}).$

Training and sampling algorithms from PFCM are available in Algorithm~\ref{alg_train} and~\ref{alg_sample}, where we have highlighted updates to consistency distillation~\cite{song2023} in blue. One notable feature is the presence of the prior image $\bm{y}$. Including, or omitting, $\bm{y}$ allows for seamless transition between the supervised (conditional generation) and unsupervised (unconditional generation) versions of PFCM. Algorithmically, we can ``omit'' the prior, or condition image, by simply setting it to empty $\bm{y}=\emptyset.$ Feeding the condition image, $\bm{y}$, as additional input to the network to directly learn a given inverse problem has been demonstrated to work well empirically for diffusion models~\cite{batzolis2021,saharia2022b,saharia2023}. This strategy has also been successful for PFGM++ and CM, as shown in recent work~\cite{hein2024}.
\begin{algorithm}
    \DontPrintSemicolon
    \SetAlgoNoLine
    \textbf{Require:} dataset $\mathcal{D},$ initial model parameter $\bm{\theta},$ learning rate $\eta$, ODE solver $\Phi(\cdot,\cdot,\textcolor{blue}{\cdot}; \bm{\phi}),$ distance metric $d(\cdot,\cdot)$, weighting function $\lambda(\cdot)$ and decay parameter $\mu$\;
    $\bm{\theta^-} \leftarrow \bm{\theta}$ \;
    \textbf{repeat} \;
    $\quad$ Sample $\textcolor{blue}{(\bm{x},\bm{y})} \sim \mathcal{D}$ and $i \sim \mathcal{U}[1,N-1]$ \;
    $\quad$ \textcolor{blue}{Set $r_{i+1}\leftarrow \sigma_{i+1} \sqrt{D}$} \;
    $\quad$ \textcolor{blue}{Sample radii $R_{i+1} \sim p_{r_{i+1}}(R)$} \;
    $\quad$ \textcolor{blue}{Sample uniform angles $\bm{v}_{i+1}\leftarrow\bm{u}_{i+1}/||\bm{u}_{i+1}||_2,$ with $\bm{u}_{i+1}\sim \mathcal{N}(\bm{0};\bm{I})$}\;
    $\quad$ \textcolor{blue}{$\bm{x}_{\sigma_{i+1}} \leftarrow \bm{x}+R_{i+1}\bm{v}_{i+1}$}\; 
    $\quad$ $\breve{\bm{x}}_{\sigma_{i}}\leftarrow \bm{x}_{\sigma_{i+1}}+(\sigma_i-\sigma_{i+1})\Phi(\bm{x}_{\sigma_{i+1}},\sigma_{i+1},\textcolor{blue}{\bm{y}};\bm{\phi})$\;
    $\quad$ $\mathcal{L}(\bm{\theta},\bm{\theta^-};\bm{\phi}) \leftarrow$\;
    $\quad \quad$ $\lambda(\sigma_i)\left(f_{\bm{\theta}}(\bm{x}_{\sigma_{i+1}},\sigma_{i+1},\textcolor{blue}{\bm{y}}),f_{\bm{{\theta^-}}}(\breve{\bm{x}}_{\sigma_{i}},\sigma_i,\textcolor{blue}{\bm{y}})\right)$\;
    $\quad$ $\bm{\theta}\leftarrow \bm{\theta}-\eta \nabla_{\bm\theta} \mathcal{L}(\bm{\theta},\bm{\theta^-};\bm{\phi})$ \;
    $\quad$ $\bm{\theta^-} \leftarrow \text{stopgrad}(\mu\bm{\theta^-}+(1-\mu)\bm{\theta})$ \;
    \textbf{until} convergence
    \caption{Training PFCM via distillation. Extended from \cite{song2023} and \cite{xu2023}.}
    \label{alg_train}
\end{algorithm}
\begin{algorithm}
    \DontPrintSemicolon
    \SetAlgoNoLine
    \textbf{Require:} PFCM $f_{\bm{\theta}}(\cdot, \cdot, \textcolor{blue}{\cdot})$ \textcolor{blue}{and condition data $\bm{y}$}\;
    \textcolor{blue}{Set $r\leftarrow \sigma_{\text{max}} \sqrt{D}$} \;
    \textcolor{blue}{Sample radii $R \sim p_r(R)$} \;
    \textcolor{blue}{Sample uniform angles $\bm{v} \leftarrow \bm{u}/||\bm{u}||_2,$ with $\bm{u}\sim \mathcal{N}(\bm{0};\bm{I})$\;
    $\bm{x} \leftarrow R\bm{v}$} \;
    $\hat{\bm{x}} \leftarrow f_{\bm{\theta}}(\bm{x}, \sigma_{\text{max}},\textcolor{blue}{\bm{y}})$ \;
    \caption{Sampling from PFCM. Extended from~\cite{song2023} and~\cite{xu2023}.}
    \label{alg_sample}
\end{algorithm}
\subsection{Task-specific sampler}
Poisson flow and diffusion models were originally introduced for unconditional image generation, where stochastic diversity is actively encouraged. This design objective, however, stands in contrast to medical inverse problems, where data fidelity is of utmost importance. As expected, given that it is fundamentally a generative model, applying Algorithm~\ref{alg_sample} yields substantial variability in the denoised CT images, consistent with results in Pfaff et al.\cite{pfaff2024} for diffusion models. To address this, we propose a specialized sampling procedure tailored to low-dose CT denoising.

Inspired by previous works, such as~\cite{pearl2023,xiang2023,chung2023,tong2024,hein2024,hein2024b}, instead of starting from a sample from the noise distribution, we propose directly inserting the noisy image $\bm{y}$, that is replace line 1-5 with $\bm{x}\leftarrow \bm{y}$, in Algorithm~\ref{alg_sample}. This, in effect, ``hijacks'' the sampling process, greatly improving data fidelity. However, this relies on the approximation $\bm{y}\approx \bm{x}_{\sigma_i}$ for some $i\in \{1,...,N\}$, that the noisy image is approximately equal to some intermediate state. For low-dose CT, especially with a relatively ``soft'' kernel, this approximation will be poor. Our key insight is that this approximation error, or mismatch~\cite{tong2024}, can effectively be mitigated by exploiting the robustness afforded by PFCM via a tunable $D.$ As it is computationally expensive to tune $D$ exhaustively, we additionally include a regularization step where we mix the denoised image with the, noisy, input image, to enhance the image quality for a given $D$. In other words, we add an additional line with $\hat{\bm{x}} \leftarrow w\hat{\bm{x}}+(1-w)\bm{x}.$ We refer to this setup as the ``task-specific'' sampler, and it is available in Algorithm \ref{alg_sample_specific}. 

\begin{algorithm}
    \DontPrintSemicolon
    \SetAlgoNoLine
    \textbf{Require:} PFCM $f_{\bm{\theta}}(\cdot, \cdot, {\cdot})$ obtained via supervised learning on the task of image denoising, noisy images $\bm{y}$, noise level $\hat{\sigma} \in \{\sigma_{\text{min}},...,\sigma_{\text{max}}\}$, and weight $w\in[0,1]$ \;
    $\bm{x}\leftarrow \bm{y}$ \; 
    $\hat{\bm{x}} \leftarrow f_{\bm{\theta}}(\bm{x}, \hat{\sigma},\bm{y})$ \;
    $\hat{\bm{x}} \leftarrow w\hat{\bm{x}}+(1-w)\bm{x}$ \;
    \caption{Task-specific sampling with PFCM.}
    \label{alg_sample_specific}
\end{algorithm}

\section{Experiments}
\subsection{Datasets}
\subsubsection{Mayo low-dose CT}
For our training and validation, we utilize the Mayo low-dose CT dataset, from the AAPM low-dose CT Grand Challenge~\cite{aapm2017}. This publicly accessible dataset encompasses imaging data from 10 patients, with images reconstructed using two distinct kernels and two different slice thicknesses. In the present study, we focus on the 1~mm slice thickness images reconstructed with the D30 (medium) kernel. The dataset is partitioned into two subsets: a training set comprising the initial 8 patients, totaling 4,800 slices, and a validation set consisting of the remaining 2 patients, encompassing 1,136 slices.

\subsubsection{Stanford low-dose PCCT}
As test data, we use clinical images from a prototype PCCT system developed by GE HealthCare, Waukesha~\cite{almqvist2024}. We include a single human subject, scanned at Stanford University, Stanford, CA under Institutional Review Board (IRB) approval. Repeat scans were used to acquire a normal dose and a low-dose image, where the low-dose image was taken with approximately a third of the dose. Due to the fact that these are repeat scans of a subject during consecutive breath holds, the slice locations will not match exactly. Hence, we cannot use quantitative metrics. Nevertheless, the repeat scans still provide an abundance of valuable qualitative information, greatly facilitating comparisons. The Protocol used to acquire the scan is available in Table~\ref{scan_table}. We reconstruct images at 40, 70, 100~keV on a $512 \times 512$ pixel grid with 0.42~mm slice thickness. This is covering the range of energy levels normally used in clinical settings, and as will be seen below, leads to large differences in noise characteristics. In addition, we reconstruct the patient using a ``sharper'' kernel, further contrasting it from our training and validation data. 
\begin{table}
    \centering
    \begin{tabular}{ccc} \toprule
	 & Effective mAs & kVp \\ \midrule 
    %ND & 290~mA & 0.990:1 & 0.5~s & 120 \\ 
    %LD & 290~mA & 1.531:1 & 0.28~s & 120 \\ 
    ND & 146.5 & 120 \\ 
    LD & 53.0 & 120 \\ 
 \bottomrule \\
    \end{tabular}
    \caption{Protocol used to acquire PCCT test data. ND and LD are normal-dose and low-dose, respectively.}
    \label{scan_table} 
\end{table}

\subsection{Implementation details}
We obtain our PFCM by first training PFGM++ and then distilling them into single-step samplers via Algorithm~\ref{alg_train}. As was shown in the original implementation of PFGM++~\cite{xu2023}, one can re-purpose training algorithms and hyperparameters for diffusion models via the alignment formula $r=\sigma \sqrt{D}$. We translate this idea to PFCM by re-purposing most hyperparameters from CM~\cite{song2023} directly. In particular, we use the setup for the LSUN $256\times 256$ experiments\footnote{As specified in ~\url{https://github.com/openai/consistency_models/blob/main/scripts/launch.sh}.} with the exception of batch size, which we reduce to $4$ and we increase to $\sigma_{\text{max}}=380$ as a large $\sigma_{\text{max}}$ was shown to be beneficial for PFGM++~\cite{xu2023} and because we have higher resolution images. $D\in (0,\infty)$ is a key hyperparameter for PFGM++ and, since PFCM is distilled PFGM++, it is also an important hyperparamter for PFCM. We use $D \in \{128, 2048, 262144\}$ for this paper. These particular values where chosen to be representative samples of small, medium and large values of $D$, relative to $N=256 \times 256.$ We limit ourselves to three values due to compute constraints. To enable efficient compute and regularization we train the network on randomly extracted $256\times 256$ patches. Random rotations and mirrorings are also applied for effective data augmentation. We surmise that the regularization of this is important to prevent overfitting since we are operating with a relatively small dataset and a very high capacity network. The networks are trained using rectified Adam~\cite{liu2019} with learning rate $1\times10^{-4}$ for 300k iterations to obtain PFGM++ and then distilled for another 300k iterations with learning rate $1\times10^{-5}$ to obtain the final PFCM. During distillation we use LPIPS~\cite{zhang2018} as metric function. The network architecture is as in~\cite{dhariwal2021} with attention layers at resolutions $[32,16,8]$, channel multiplier $256$, and channels per resolution $[1, 1, 2, 2, 4, 4].$ In order to provide further regularization, we set dropout to $10\%$ when training PFGM++. 

The task-specific sampling introduces two additional hyperparameters: $w \in [0,1]$ and $i \in \{1,...,N\}$. We conducted a grid search over $i\in \{30,31,...,39,40\}$ and $w \in \{0.5,0.6,...,0.9,1.0\}$ for each $D \in  \{128,2048,262144\}$ as well as for the consistency model. The boundaries of the search grid were set heuristically based on initial experiments to strike a balance between performance and compute requirements. Best results are defined as lowest Learned Perceptual Image Patch Similarity (LPIPS~\cite{zhang2018}) on the low-dose CT validation set. The optimal hyperparameters for the task-specific sampler are presented in Table~\ref{hyper_table}. 

Notably, as will be seen in the results section, this task-specific sampling relies heavily on the added robustness to missteps in the solution trajectory of PFCM compared to CM. We surmise that similar task-specific sampling algorithms can favorably be employed to solve a range of different inverse problems. In addition, such task-specific algorithms can be used to solve different inverse problems using a single pre-trained PFCM in the unsupervised setting. Due to compute resource constraints, we leave both of these avenues to future work.
\begin{table}
    \centering
    \begin{tabular}{ccccc} \toprule
	 $D$ & 128 & 2048 & 262144 & CM~\cite{song2023}  \\ \midrule 
    $i$ & 30 & 31 & 33 & 34\\ 
    $w$ & 0.7 & 0.6 & 0.5 & 0.5\\ 
 \bottomrule \\
    \end{tabular}
    \caption{Hyperparameters for task-specific sampler.}
    \label{hyper_table} 
\end{table}

\subsection{Comparison to other methods}
We compare PFCM against several popular benchmarks in the low-dose CT literature, BM3D~\cite{makinen2020}, RED-CNN~\cite{chen2017}, WGAN-VGG~\cite{yang2018}, in addition to state-of-the-art deep generative models, diffusion models (EDM)~\cite{karras2022}, PFGM++~\cite{xu2023}, and CM~\cite{song2023}. BM3D is a popular non-deep learning based denoising technique, shown to be the best performing alternative to deep learning based methods in~\cite{chen2017}. We use a Python implementation\footnote{\url{https://pypi.org/project/bm3d/}.} of BM3D and $\sigma_{\text{BM3D}}$ was set to the standard deviation of a flat region-of-interest (ROI) measured in the low-dose CT validation data. RED-CNN was trained on extracted $55 \times 55$ overlapping patches from the training data. WGAN-VGG was trained on randomly extracted $64\times64$ patches and everything else set as in the original implementation~\cite{yang2018}. EDM and PFGM++ use the same hyperparameters, except for $D$, and are trained as detailed above. Finally, the CM trained using the same setup as for PFCM, also detailed above. 

\subsection{Evaluation methods}
For quantitative assessment we employ standard metrics in the low-dose CT denoising literature such as the peak signal-to-noise ratio (PSNR) and structural similarity index (SSIM~\cite{wang2004}). However, it is well known that these metrics do not necessarily reflect human perception~\cite{zhang2018}. This is particularly evident when it comes to over-smoothing as PSNR simply inversely proportional to the mean squared error (MSE). This pixel-wise comparison is at odds with how human perception works. Whilst the underlying mechanics of human perception are very complex, it has been shown that metrics based on extracted features from pre-trained convolutions neural networks, such as LPIPS, correlate well with human perception~\cite{zhang2018}. Another way to think of LPIPS is in light of the perception-distortion trade-off. LPIPS is a distortion metric; however, it is ``more perceptual'' than, for instance, PSNR and therefore results in a weaker trade-off~\cite{blau2018}. We use a version of LPIPS with AlexNet~\cite{krizhevsky2014} as feature extractor from the official repository\footnote{\url{https://github.com/richzhang/PerceptualSimilarity}.}. The fact that AlexNet was trained on natural RGB images presents two challenges. First, we need to force the network to accepted grayscale instead of RGB images. We follow the standard approach of simply feeding triplets of the repeated grayscale image. Second, and more importantly, this pre-trained network to CT images means that we are using it out-of-distribution. With that caveat stated, we observe that LPIPS works very well on average. However, we take all considered metrics into account. We also report the number of function evaluations (NFEs) required during inference. In other words, the total number of passes through the network to generate the desired denoised image, allowing for an assessment of computational requirements. 

\subsection{Results}
\subsubsection{Ablation study of task-specific sampler}
\begin{figure*}
    \centering
    \includegraphics[width=0.9\linewidth]{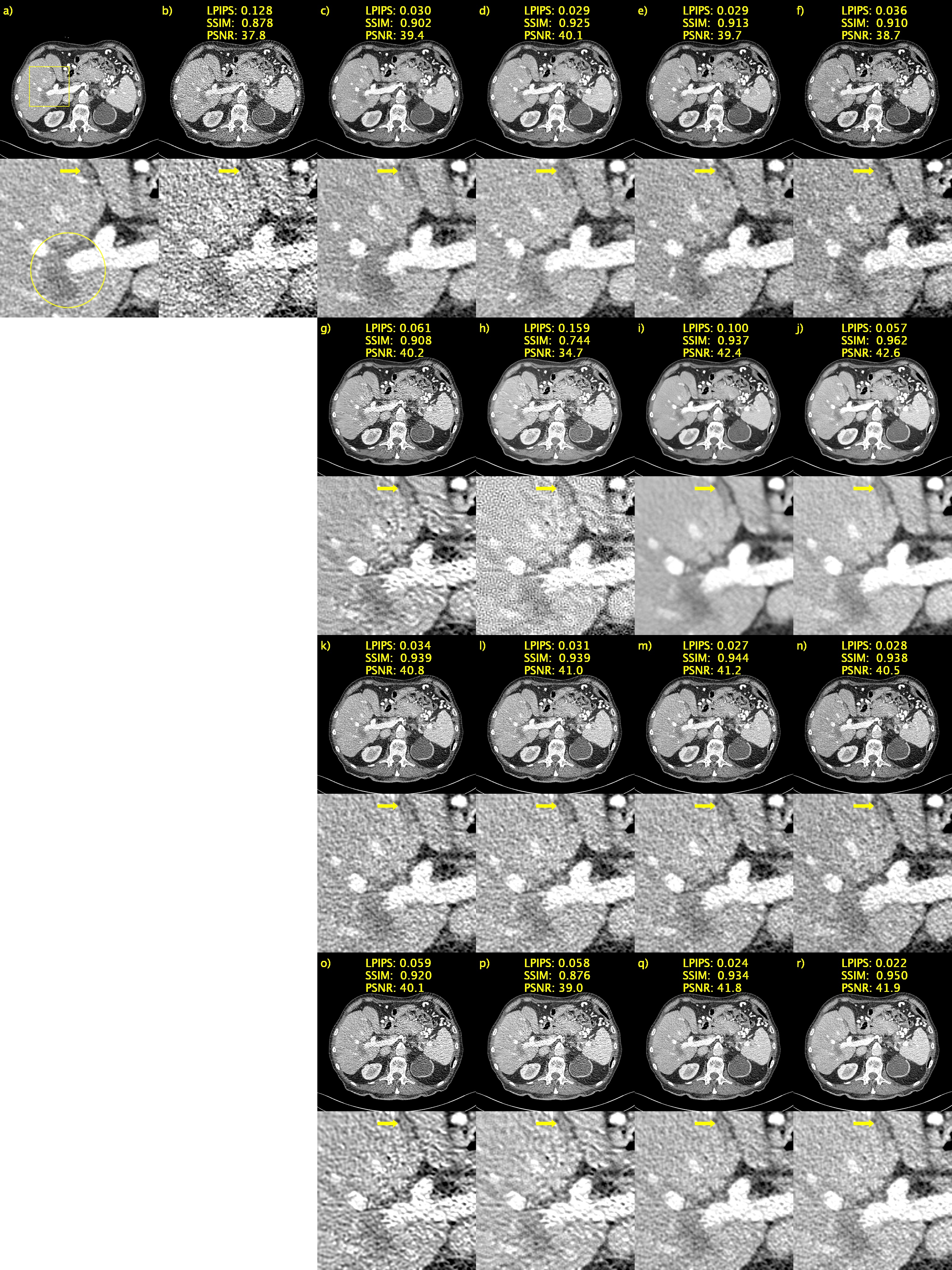}
    \caption{Ablation study of task-specific sampler: hijacking (h) + regularization (r). a) NDCT, b) LDCT, c) CM~\cite{song2023}, d) PFCM-262144, e) PFCM-2048, f) PFCM-128, g) CM+h, h) PFCM-262144+h, i) PFCM-2048+h, j) PFCM-128+h, k) CM+r, l) PFCM-262144+r, m) PFCM-2048+r, n) PFCM-128+r, o) CM+h+r, p) PFCM-262144+h+r, q) PFCM-2048+h+r, r) PFCM=128+h+r. Yellow circle added to emphasize lesion. Yellow arrow placed to emphasize detail. 1~mm-slices. Window setting [-160,240]~HU.}
    \label{18_ablation}
\end{figure*}

We start with an ablation study of the task-specific sampler. This sampler has two parts: ``hijacking'' and ``regularization''. Qualitative results, along with LPIPS, SSIM, and PNSR for the particular slice considered, are available in Fig.~\ref{18_ablation}. As PFCM is a distilled version of PFGM++, it inherits the ability to trade off robustness for rigidity by tuning $D\in (0,\infty)$. To shed light on this key hyperparameter, we include results for $D\in \{128, 2048, 262144\}$, in addition to CM, which is a distilled version of a diffusion model, the limiting case where $D\rightarrow \infty.$ As specified above in Table~\ref{hyper_table}, the hyperparameters for each are optimized separately. We use a ``$+$'' notation to illustrate that we append hijacking, regularization, or both. The top row, c)-f), shows the results from the vanilla formulation in Algorithm~\ref{alg_sample}, with no hijacking nor regularization, for $D\in \{128, 2048, 262144\}$ and $D\rightarrow \infty$ (CM). Although perception is good, as the noise characteristics closely resembles that of the NDCT image, shown in a), there is significant distortion, as emphasized by the yellow arrow. Moreover, the visual quality is superior for larger $D.$ The second row, g)-j), shows the results where we only apply hijacking, resulting in a ``mismatch'' as the low-dose CT image is a poor approximation of underlying intermediate states of the diffusion and Poisson flow processes. Consistent with this, the results for the CM and $D=262144$, shown in g) and h), are very poor. Our key insight is that we can leverage the robustness of PFCM for relatively small $D$ to mitigate issues resulting from this mismatch. This is clearly demonstrated in i) and j), showing results employing only hijacking for $D=2048$ and $D=128$, respectively. For $D=2048$, however, we do observe over-smoothing. This seems to be significantly less prominent for $D=128$, although still slightly evident. We surmise that further tuning of $D$, as well as the number of noise scales, may lead to excellent results with only hijacking. However, due to limited compute resources, we opt instead to include an additional regularization step. Results for only this regularization step, which is just a weighted average between the denoised and noisy image, are available on the third row, k)-n). Finally, in the last row, we show the results for our full task-specific sampler, employing both hijacking and regularization, in o)-r). The mismatch from hijacking is still very prominent for CM and $D=262144$, shown in o) and p), respectively, leading to poor image quality. Setting a lower $D$, yielding superior robustness, effectively handles this mismatch, and together with regularization, yields excellent perceptual quality, and data fidelity, as shown in q) and r) for $D=2048$ and $D=128$, respectively. For both e) and f), compared to q) and r), all metrics improve. Crucially, this is due to the combination of our task-specific sampler, and the robustness afforded by setting a small $D$ in PFCM. 

Quantitative results, on the entire validation set, from this ablation study are available in Table~\ref{quant_ablation}. These results closely align with the qualitative analysis in Fig.~\ref{18_ablation}, robustness to hijacking greatly improves with a smaller $D$. We note, however, that there is seemingly a slight trade-off between LPIPS and PSNR, as adding the additional regularization greatly improves LPIPS at a cost of a marginal decrease in PSNR for $D=128.$ Comparing with PFGM++, we can see that the net performance gain of PFCM with the task-specific sampler is 79 $\times$ faster sampling than PFGM++~\cite{xu2023} and EDM~\cite{karras2022} while performing at least on-par in terms of LPIPS, SSIM, and PSNR. 
\begin{table*}
\centering
\begin{tabular}{cccccc} \toprule
    & D & LPIPS ($\downarrow$) & SSIM ($\uparrow$)&	PSNR ($\uparrow$) &	NFE ($\downarrow$) \\ \midrule 
    EDM~\cite{karras2022} & & $0.012 \pm 0.005$	& $0.97	\pm 0.01$	& $43.8	\pm 1.23$ & 79\\
    CM~\cite{song2023} & &$0.015 \pm 0.005$& $ 0.94\pm 0.01$ & $42.0\pm 0.83 $  & 1\\ 
    +hijacking & & $0.027 \pm 0.012$& $ 0.94\pm 0.01$ & $42.7\pm 0.82 $  & 1\\  
    +regularization & & $0.016 \pm 0.006$& $ 0.97\pm 0.01$ & $44.0\pm 1.16 $  & 1\\  
    +hijacking and regularization & & $0.025 \pm 0.010$& $ 0.96\pm 0.02$ & $44.4\pm 1.69 $  & 1\\  \midrule 
    PFGM++~\cite{xu2023} & 128 &  $0.012\pm	0.005$	& $0.96\pm	0.01$	& $43.5\pm 1.22$ & 79\\
     & 2048 &  $0.012\pm	0.005$	& $0.97\pm	0.01$	& $43.7\pm 1.24$ & 79\\
     & 262144 &  $0.012\pm	0.005$	& $0.96\pm	0.01$	& $43.6\pm 1.15$ & 79\\
    PFCM &128 &$0.018 \pm 0.005$& $0.94 \pm 0.01 $ & $41.8\pm 0.88$ & 1\\ 
     &2048 &$0.018 \pm 0.004$& $0.91 \pm 0.01 $ & $41.3\pm 0.82$ & 1\\ 
     &262144  &$0.015 \pm 0.004$& $0.96 \pm 0.01 $ & $42.4\pm 0.81$ & 1\\ 
    +hijacking & 128 &$ 0.031 \pm 0.011$& $\bm{0.98} \pm 0.01 $ & $\bm{45.7}\pm 1.10$ & 1 \\ 
    & 2048 &$0.054 \pm 0.021$& $0.96 \pm 0.01 $ & $45.4\pm 1.07$ & 1\\
    & 262144 &$ 0.196 \pm 0.022 $& $0.74 \pm 0.02 $ & $35.3\pm 0.33$ & 1 \\
    +regularization & 128 &$0.014 \pm 0.004$& $0.97 \pm 0.01 $ & $43.6 \pm 0.98$ & 1 \\ 
    & 2048 &$0.013 \pm 0.004$& $0.96 \pm 0.01 $ & $43.6\pm 0.96$ & 1\\
    & 262144 &$ 0.016 \pm 0.006 $& $0.97 \pm 0.01 $ & $44.1\pm 1.19$ & 1 \\
    +hijacking + regularization & 128 &$\bm{0.010} \pm 0.004$& $\bm{0.98} \pm 0.01 $ & $45.3\pm 1.28$ & 1 \\ 
    & 2048 &$\bm{0.010} \pm 0.004$& $0.97 \pm 0.01 $ & $45.3 \pm 1.32$ & 1\\
    & 262144 &$ 0.024 \pm 0.011 $& $0.90 \pm 0.01 $ & $41.0\pm 0.70$ & 1 \\
\bottomrule \\
\end{tabular}
\caption{Ablation study of task-specific sampler: hijacking + regularization. Mean and standard deviation of LPIPS, SSIM, and PSNR in the low-dose CT validation set. $\downarrow$ means lower is better. $\uparrow$ means higher is better. Best results in bold.}
\label{quant_ablation} 
\end{table*}

\subsubsection{Mayo low-dose CT}
\begin{figure}
    \centering
    \includegraphics[width=\columnwidth]{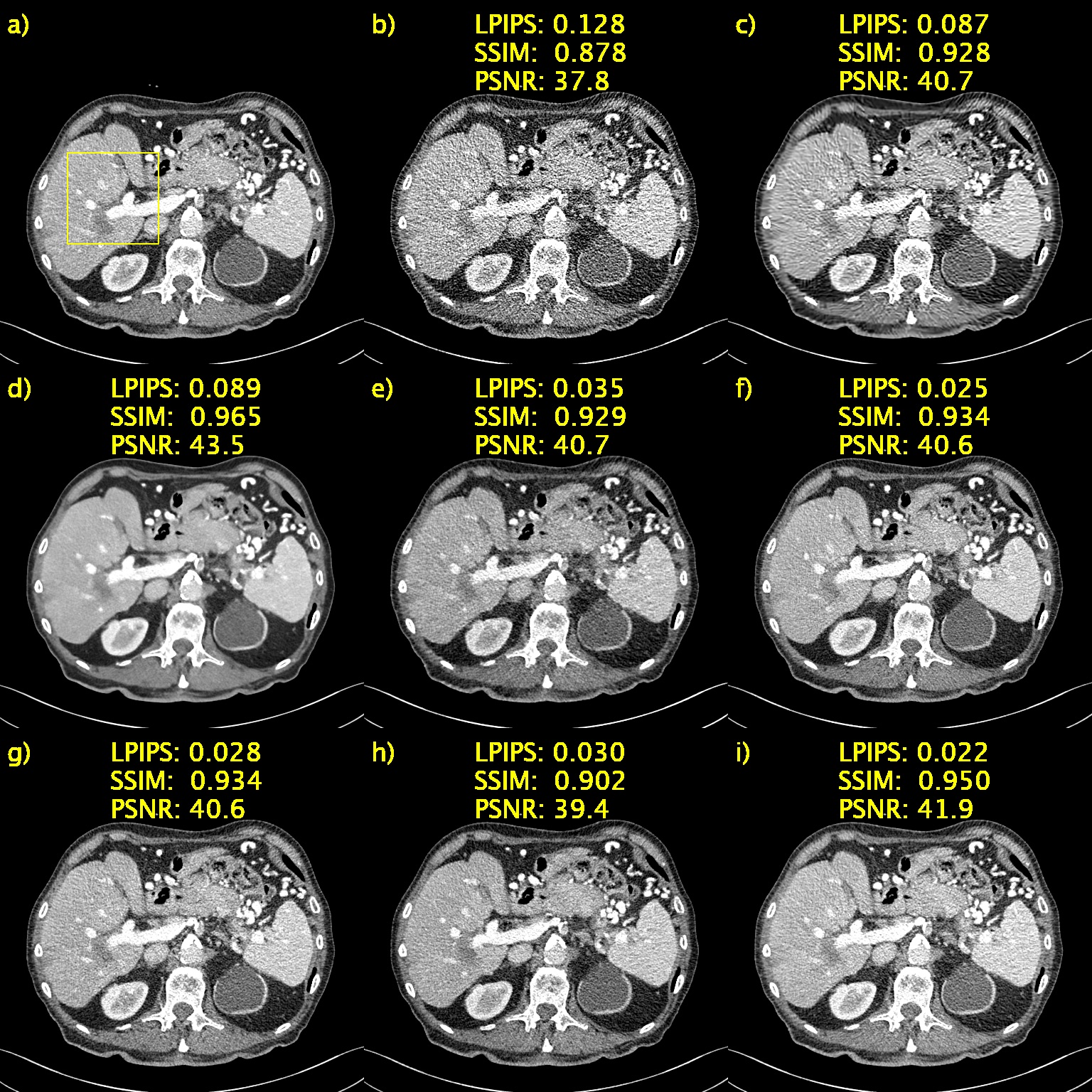}
    \caption{Results on the Mayo low-dose CT validation data. Abdomen image with a metastasis in the liver. a) NDCT, b) LDCT, c) BM3D~\cite{makinen2020}, d) RED-CNN~\cite{chen2017}, e) WGAN-VGG~\cite{yang2018}, f) EDM~\cite{karras2022}, g) PFGM++~\cite{xu2023}, h) CM~\cite{song2023}. i) Proposed. Yellow box indicating ROI shown in Fig. \ref{18_ex}. 1~mm-slices. Window setting [-160,240]~HU.}
    \label{18}
\end{figure}

\begin{figure}
    \centering
    \includegraphics[width=\columnwidth]{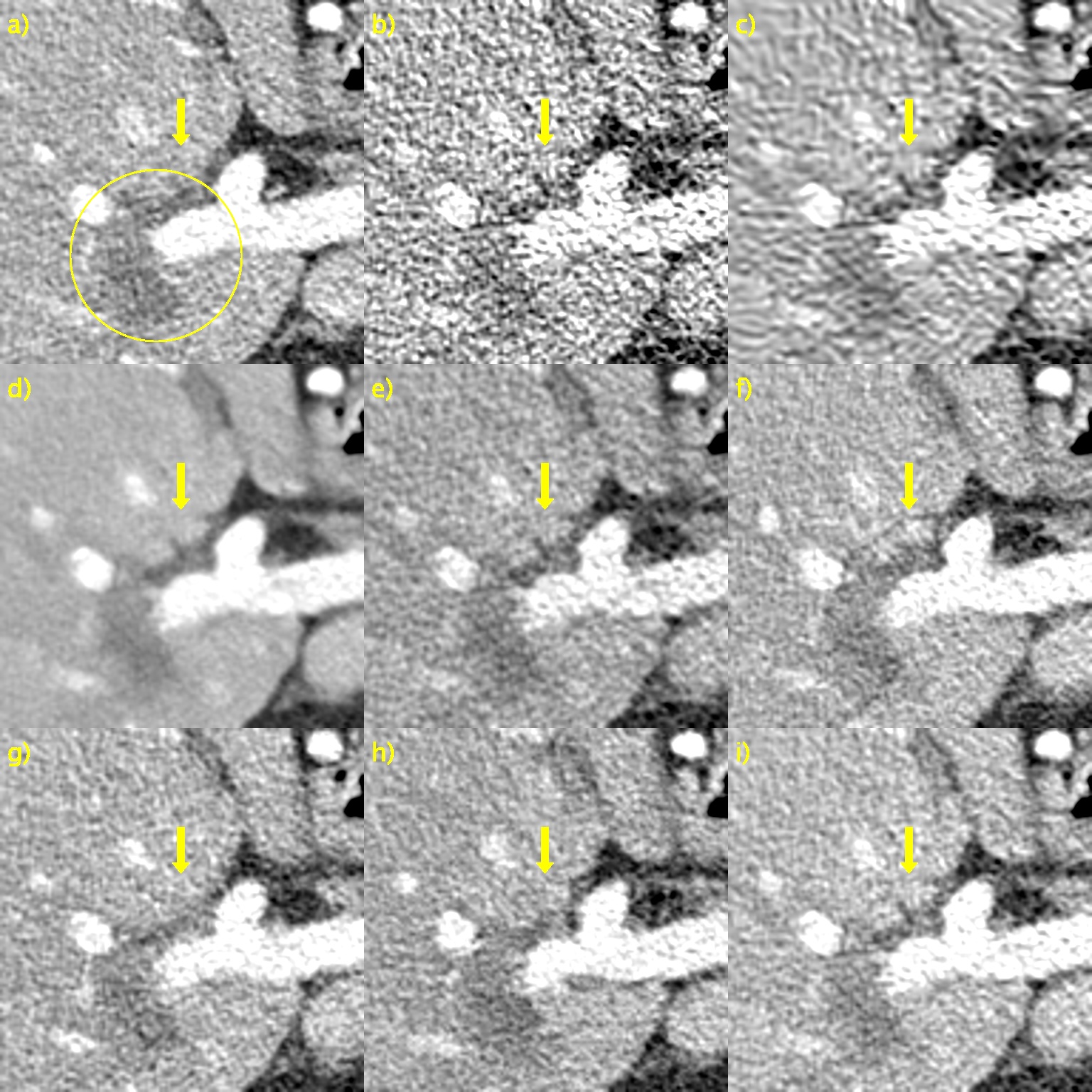}
    \caption{ROI in Fig. \ref{18} magnified to emphasize details. a) NDCT, b) LDCT, c) BM3D~\cite{makinen2020}, d) RED-CNN~\cite{chen2017}, e) WGAN-VGG~\cite{yang2018}, f) EDM~\cite{karras2022}, g) PFGM++~\cite{xu2023}, h) CM~\cite{song2023}, i) Proposed. Yellow circle added to emphasize lesion. Yellow arrow placed to emphasize detail. 1~mm-slices. Window setting [-160,240]~HU.}
    \label{18_ex}
\end{figure}

We now move to our quantitative and qualitative results on the Mayo low-dose CT validation set. Qualitative results are shown in Fig.~\ref{18} and~\ref{18_ex}; LPIPS, SSIM, and PSNR are also reported for this particular slice. This patient has a lesion in the liver, which we have emphasized with a yellow circle. We have additionally place a yellow arrow on a detail that exhibits significant stochastic variation. To ease comparisons we only include the top performing PFCM with task-specific sampler, achieved for $D=128$. In addition, we only include the best PFGM++, which as was achieved for $D=2048.$ BM3D~\cite{makinen2020}, in c), yields an image with strange noise characteristics. However, it does a good job keeping salient details visible in the LDCT image in b) intact. RED-CNN~\cite{chen2017}, shown in d), produces an image that is visually very pleasing. However, the target is the NDCT image, shown in a), and thus the denoising in RED-CNN is significantly over-shooting the target, resulting in over-smoothing. This is fully expected since RED-CNN is trained with a simple MSE loss. To address this issue, WGAN-VGG~\cite{yang2018} uses a combination of an adversarial and perceptual loss to obtain a denoised image with similar noise characteristics and the NDCT image. Comparing e) with a), we can see that WGAN-VGG achieves this objective. EDM~\cite{karras2022}, results in a reasonable sample from the desired posterior at first glance. However, on closer inspection we can see details that are distinct from both the NDCT and LDCT images. We have indicated one such ``hallucination'' with a yellow arrow. There is a hint of this ``lesion'' in this NDCT image, and we can see it in c), d), e), and j). However, for EDM, this detail has been ``misclassified'' as signal and thus been enhanced, resulting in a detail that the radiologist might interpret as a lesion. Such stochastic variation might not be a big problem is general computer vision task, but can have dramatic consequence in medical imaging. PFGM++~\cite{xu2023}, shown in g), does not exhibit this particular detail but there is still a large degree of undesirable stochastic variation. This behavior is also noted for CM~\cite{song2023}, shown in h). This is illustrating the need to ``styme'' the generative process in order to ensure better data fidelity when adapting diffusion-style models to inverse problem solving in medical imaging. This is what is done with our proposed method, PFCM with the task-specific sampler, available in j). As also seen in Fig.~\ref{18_ablation}, the proposed method results in an image with noise characteristics closely matching that of the NDCT image, shown in a), with all salient details in the NDCT image, shown in b), remaining intact. 

\begin{figure}
    \centering
    \includegraphics[width=\columnwidth]{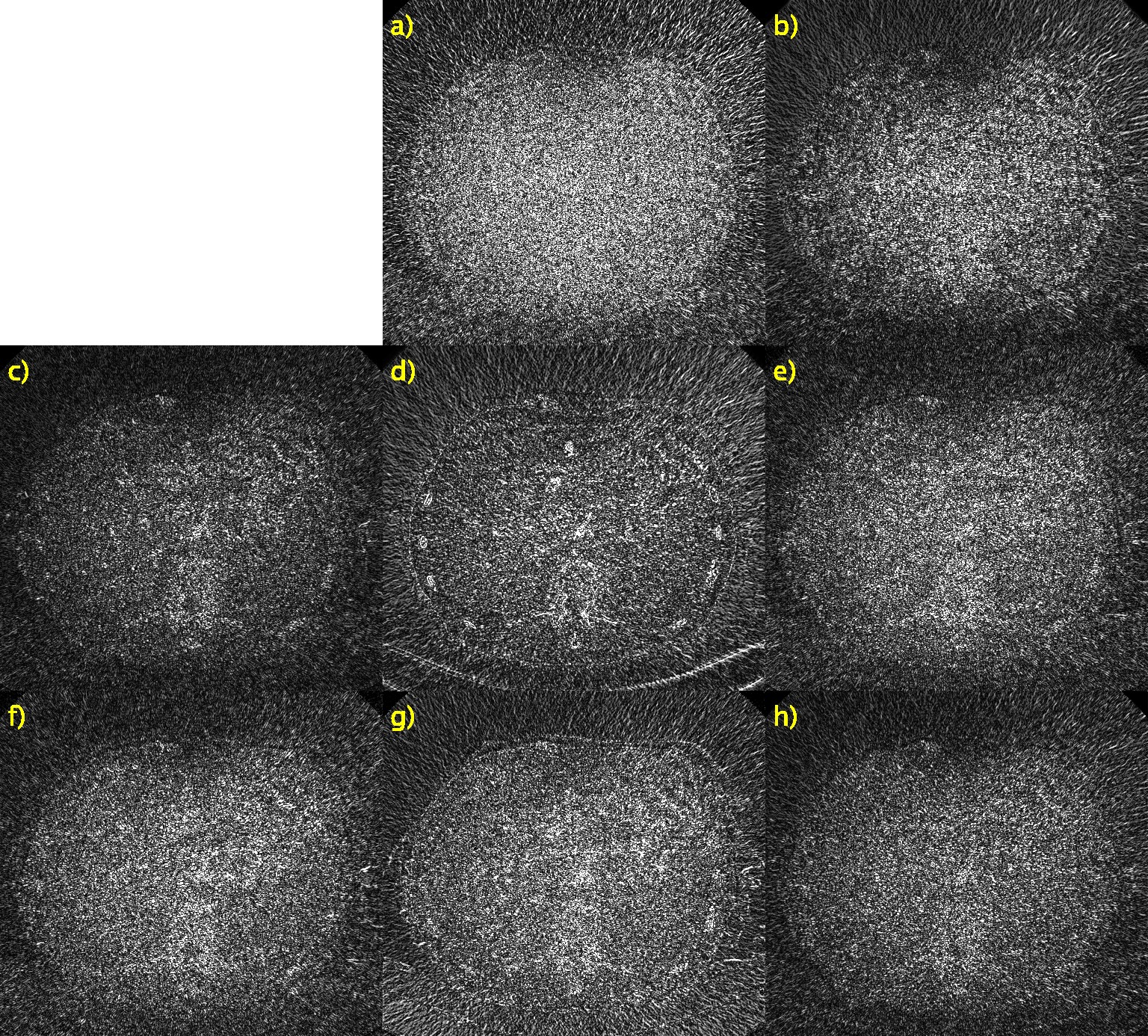}
    \caption{Absolute difference with respect to the NDCT image. a) LDCT, b) BM3D~\cite{makinen2020}, c) RED-CNN~\cite{chen2017}, d) WGAN-VGG~\cite{yang2018}, e) EDM~\cite{karras2022}, f) PFGM++~\cite{xu2023}, g) CM~\cite{song2023}, h) Proposed. 1~mm-slices. Window setting [0,100]~HU.}
    \label{18_diff}
\end{figure}

To further validate our results, we include the absolute difference, with respect to the NDCT image, for each method in Fig.~\ref{18_diff}. In this difference image, we strive to see only noise, as is the case for the LDCT image shown in a). A hint of the spine is visible in all methods. In addition, this analysis shows a significant limitation of WGAN-VGG, as there are many structures clearly discernible. The proposed method, shown in h), on the other hand, results in a very ``clean'' difference map. 

\begin{figure}
    \centering
    \includegraphics[width=\columnwidth]{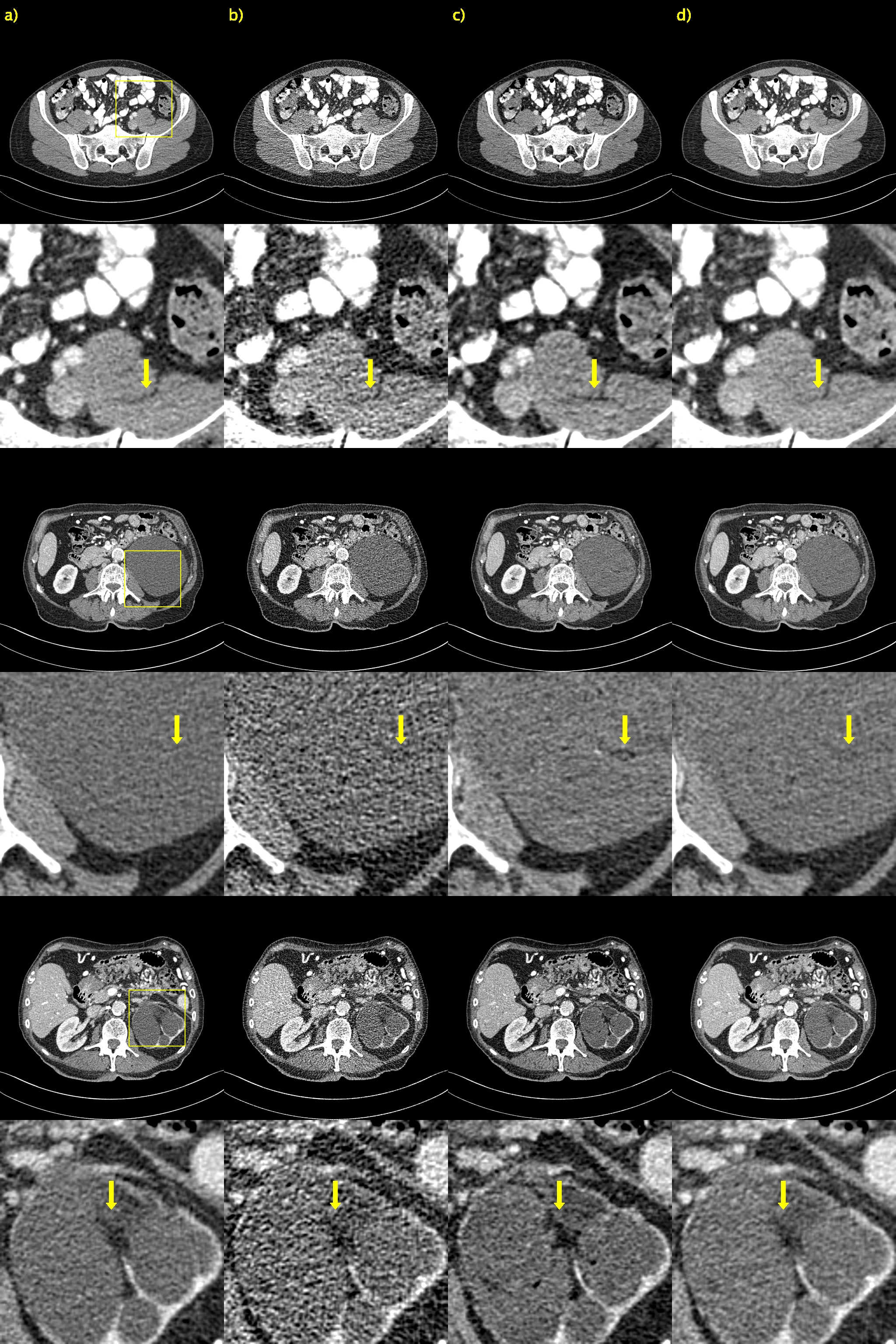}
    \caption{Additional results for task-specific sampler. a) NDCT, b) LDCT, c) PFCM-128, d) PFCM-128+task-specific sampler. Yellow arrow placed to emphasize detail. 1~mm-slices. Window setting [-160,240]~HU.}
    \label{extra}
\end{figure}

Finally, to emphasize the importance of the task-specific sampler, and to further illustrate its effectiveness in improving data fidelity and dealing with ``hallucinations'', we have include three additional slices in Fig.~\ref{extra}. We have placed yellow arrows on hallucinations produced by PFCM, where sample diversity is encouraged, that are effectively dealt with using the task-specific sampler. 

Quantitative results are available in Table~\ref{quant_table}. RED-CNN performs very well according to SSIM and PSNR. However, recall that PNSR is inversely proportional to the MSE loss and thus this does not penalize over-smoothing. As was seen in Fig.~\ref{18} and~\ref{18_ex}, RED-CNN results in a significantly over-smoothed image. This is, importantly, penalized heavily be LPIPS.  WGAN-VGG, on the other hand, performs very well in terms of LPIPS, consistent with the qualitative results. PFCM, our proposed method, is the top performer in terms of LPIPS and performs on-par with RED-CNN according to SSIM. Notably, PFCM significantly outperforms CM in terms of all metrics considered, emphasizing the benefit of our suggest approach. 

\begin{table*}
\centering
\begin{tabular}{ccccc} \toprule
    & LPIPS ($\downarrow$) & SSIM ($\uparrow$)&	PSNR ($\uparrow$) &	NFE ($\downarrow$)\\ \midrule 
    LDCT   & $0.075 \pm 0.022$ & $0.94 \pm 0.02 $ & $41.5\pm 1.60$ & \\
    BM3D~\cite{makinen2020} & $0.050 \pm 0.014$ & $0.97 \pm 0.01 $ & $45.0\pm 1.63$ &\\ 
    RED-CNN~\cite{chen2017} & $0.048 \pm 0.018$ & $\bm{0.98} \pm 0.06 $ & $\bm{46.8}\pm 1.21$ & 1\\ 
    WGAN-VGG~\cite{yang2018} & $0.019 \pm 0.005$ & $0.96 \pm 0.01 $ & $43.2\pm 0.93$ & 1\\  
    EDM~\cite{karras2022} &$0.012 \pm 0.005$ & $0.97 \pm 0.01 $ & $43.8\pm 1.23$ &  79 \\ 
    PFGM++~\cite{xu2023} &$0.012 \pm 0.005$& $0.97 \pm 0.01 $ & $43.7\pm 1.24$ &  79\\ 
    CM~\cite{song2023} &$0.015 \pm 0.004$& $ 0.96\pm 0.01$ & $42.4\pm 0.81 $ &  1 \\ 
    PFCM+task-specific sampler &$\bm{0.010} \pm 0.004$& $\bm{0.98} \pm 0.01 $ & $45.3\pm 1.28$ &  1\\ 
\bottomrule \\
\end{tabular}
\caption{Quantitative results. Mean and standard deviation of LPIPS, SSIM, and PSNR in the low-dose CT validation set. $\downarrow$ means lower is better. $\uparrow$ means higher is better. Best results in bold.}
\label{quant_table} 
\end{table*}

\subsubsection{Stanford low-dose PCCT}
\begin{figure}
    \centering
    \includegraphics[width=0.95\columnwidth]{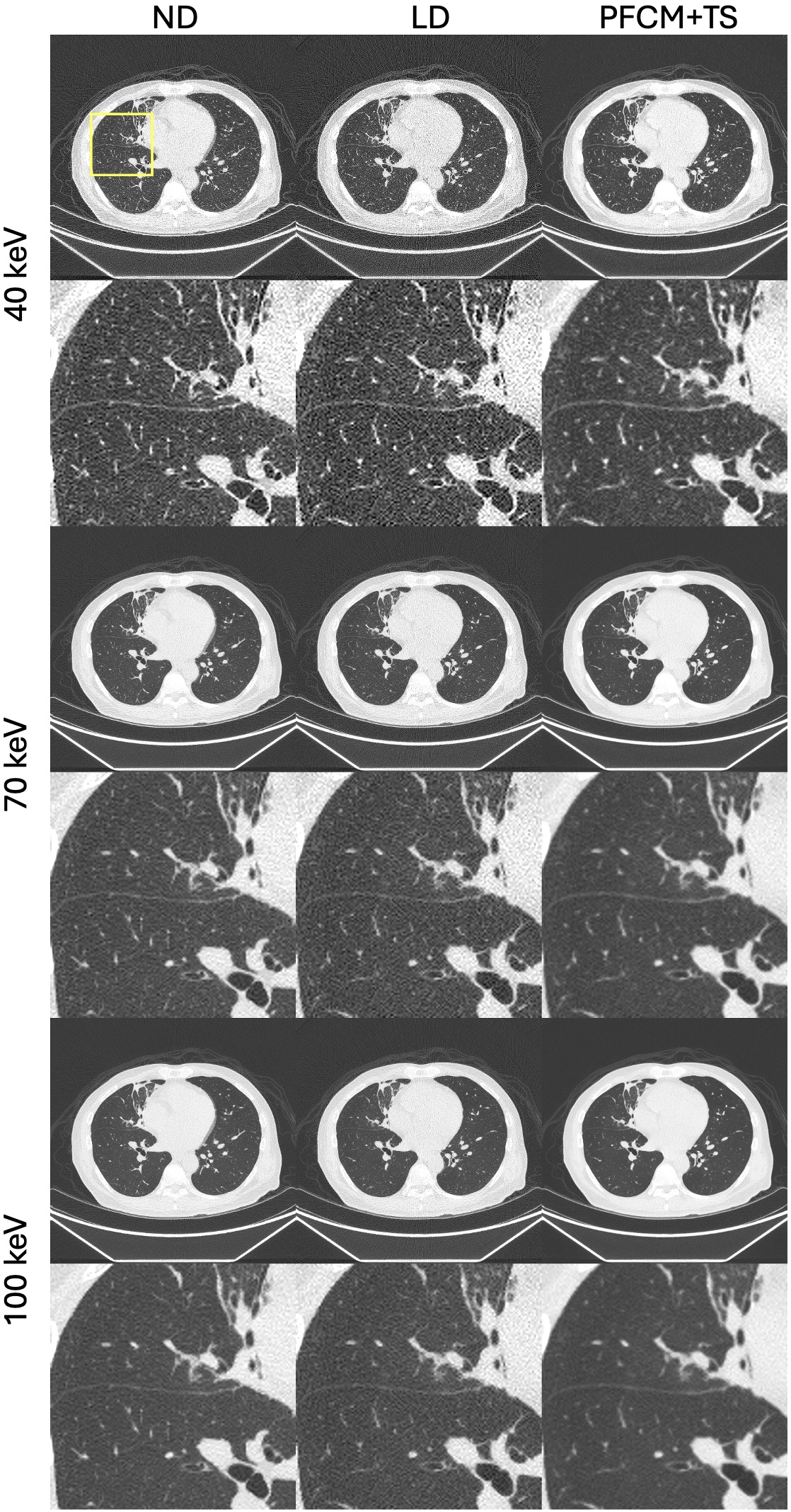}
    \caption{Results on the PCCT test data. ND and LD are normal-dose and low-dose, respectively. PFCM+TS is the proposed method. Yellow box indicating magnified ROI. 0.42~mm-slices. Window setting [-1350,150]~HU.}
    \label{non_contrast}
\end{figure}

Finally, qualitative results on the test data are available in Fig.~\ref{non_contrast}. From left to right, we show the normal-dose (ND), low-dose (LD), and processed image (PFCM+TS). From top to bottom we have stacked to rows of original image and magnified ROI for 40, 70, and 100~keV. Since the ND and LD images are repeat scans of a human subject, the slice locations will not match exactly. This is a thorax scan and we have set the window level and width accordingly. We can see that our proposed method is capable of suppressing noise whilst keeping details intact at all energy levels. We overshoot the target in the ND image due to the fact that it is trained on data where the LD image has a $4\times$ lower dose whereas here the difference is only a factor of $2.8\times$.

\section{Discussion and Conclusion}
In this paper, we introduced Poisson Flow Consistency Models (PFCM), a novel family of deep generative models that successfully combines the robustness of PFGM++ with the efficient single-step sampling of CM. By generalizing consistency distillation to PFGM++, through a simple change-of-variables and an updated noise distribution, we have created a framework that inherits the ability to trade off robustness for rigidity via the hyperparameter $D$. This flexibility allowed us to find an optimal balance that outperforms traditional CM on the Mayo low-dose CT dataset. To boost performance further and to adapt this deep generative model to medical imaging, we proposed a ``task-specific'' sampler that exploits the robustness of PFCM. By further conditioning on the low-dose CT image, we achieved significant improvements in data fidelity, leading to excellent performance across all evaluated metrics. Moreover, the successful application of our method to clinical images from a prototype photon-counting system, reconstructed using a sharper kernel and at various energy levels, demonstrates versatility and robustness.

Future work could focus on further optimizing the task-specific sampler, tuning of $D$, and exploring the application of PFCM to other inverse problems. Notably, the current framework is based on supervised learning. However, PFCM extends to the unsupervised case as well, enabling single or multi-step sampling for unconditional image generation, in addition to the possibility of adapting one single trained network to a range of different inverse problems.

\section*{Acknowledgment}
D. H. discloses research collaboration with, and consultancy for, GE HealthCare. G.S. is an employee of GE HealthCare. A. W. discloses research support from GE HealthCare. G. W. discloses research support from GE HealthCare. 

\bibliographystyle{IEEEtran}
\bibliography{ref}
\end{document}